\documentclass[prd,twocolumn,showpacs,preprintnumbers,floats,floatfix,apsrev,amsmath,amsfonts,nofootinbib]{revtex4-1}
\usepackage{color}
\usepackage{graphicx}
\usepackage{dcolumn}
\usepackage{txfonts}

\usepackage{natbib}
\usepackage{hyperref}
\usepackage{hypernat}
\newtheorem{thm}{Theorem}
\newtheorem{df}{Definition}

\begin{document}

\title{Bekenstein Inequalities and Nonlinear Electrodynamics}
\author{M.~L.~Pe\~{n}afiel}\email{mpenafiel@cbpf.br}
\affiliation{CBPF - Centro Brasileiro de
Pesquisas F\'{\i}sicas, Xavier Sigaud st. 150,
zip 22290-180, Rio de Janeiro, Brazil.}

\author{F.~T.~Falciano}\email{ftovar@cbpf.br}
\affiliation{CBPF - Centro Brasileiro de
Pesquisas F\'{\i}sicas, Xavier Sigaud st. 150,
zip 22290-180, Rio de Janeiro, Brazil.}

\date{\today}

\begin{abstract}

Bekenstein and Mayo proposed a generalised bound for the entropy, which implies some inequalities between the charge, energy, angular momentum, and the size of the macroscopic system. Dain has shown that Maxwell's electrodynamics satisfies all three inequalities. We investigate the validity of these relations in the context of nonlinear electrodynamics and show that Born-Infeld electrodynamics satisfies all of them. However, contrary to the linear theory, there is no rigidity statement in Born-Infeld. We study the physical meaning and the relationship between these inequalities and, in particular, we analyse the connection between the energy-angular momentum inequality and causality. 
\end{abstract}


\maketitle

\section{Introduction}

Bekenstein bounds and inequalities constitute a set of universal relations between physical quantities and fundamental constants of nature~\citep{Bekenstein1973,Bek81}. They were initially formulated from gedanken experiments within the scope of black hole thermodynamics (BHT), which is a formal analogy between gravitational compact systems and the three laws of thermodynamics~\citep{Wald2001, Szabados2009, Dain2012, Dain2014}. This formalism is a sound effort to reconcile thermodynamics and black hole physics, an example of which is the Generalized Second Law (GSL)~\citep{Bekenstein1973, Bekenstein1972, Bekenstein1974}. The Bekenstein bounds and inequalities can be seen as necessary conditions in order to guarantee GSL and the consistency of General Relativity with the laws of Thermodynamics.

However, since its first proposal, there has been numerous generalisations of Bekenstein inequalities~\citep{Zaslavskii1992,Hod99}. General arguments seem to point to a consensus of the existence but there are still controversies on their precise formulation. The most general inequality was obtained by Bekenstein and Mayo in \citep{Bek1999} that relates the entropy of a system with domain $\Sigma$ with the size, energy, angular momentum and charge as
\begin{equation} \label{eq:BekMayo}
\frac{\hbar c}{2\pi \kappa_{B}}S\le\sqrt{\left(\mathcal{ER}\right)^2-c^2J^2}-{{Q^2\over{8\pi}}} \ .
\end{equation}

In the relation above, $\mathcal{R}$ is defined as the radius of the minimum sphere, $\mathcal{B_R}$, that circumscribes the domain $\Sigma$. It can be shown that inequality \eqref{eq:BekMayo} is saturated for the case of a Kerr-Newman black hole~\citep{Bek1999}. This result comes with no surprise as long as the inequalities were constructed within BHT. Notwithstanding, it also shows that one should expect equality to always be reached in the most symmetric configuration. In addition, since the entropy of a system is always nonnegative, eq.~(\ref{eq:BekMayo}) also implies 
\begin{equation} \label{eq:BekC}
\mathcal{E}^2\ge\frac{Q^4}{64\pi^2\mathcal{R}^2}+\frac{c^2J^2}{\mathcal{R}^2} \ ,
\end{equation}
where equality happens if $S=0$. Contrary to the first inequality, the only fundamental constant appearing in \eqref{eq:BekC} is the speed of light, which makes theories of electrodynamics particularly appropriate to test it.
Along this line, Dain \citep{Dain2015} has proved that the above inequality holds for any field configuration of Maxwell electrodynamics.

We can still decompose \eqref{eq:BekC} in two particular cases: one for vanishing angular momentum and the other for neutral objects. For vanishing angular momentum, $J=0$, the energy and charge of a system have to satisfy the inequality 
\begin{equation} \label{eq:BekQ}
\mathcal{E}\ge{Q^2\over{8\pi\mathcal{R}}}\ . 
\end{equation}

The equality in this case states that the total energy of the system equals the electrostatic energy of a spherical thin shell of radius $\mathcal{R}$ and constant surface charge density in Maxwell's theory. Thus, the equality is associated with the most symmetric case in the linear electrodynamics theory.

For neutral objects, $Q=0$, we obtain a quasi-local inequality that relates the energy of the electromagnetic field and its angular momentum for the region $\Sigma$ as
\begin{equation} \label{eq:BekJ}
\mathcal{E}(\Sigma)\ge{c\, \vert J(\Sigma)\vert\over{\mathcal{R}}} \ .
\end{equation}

For Maxwell electrodynamics, the total energy $\mathcal{E}$ is always greater than $\mathcal{E}(\Sigma)$ and hence inequality \eqref{eq:BekJ} implies \eqref{eq:BekC} with $Q=0$. However, there is no such guarantee for nonlinear electrodynamics. Besides, there seems to have no straightforward interpretation for inequality \eqref{eq:BekJ}. We can gain some insight by looking again to the case of a rigid slowly rotating spherical thin shell in Maxwell electrodynamics. Within this approximation, it can be shown that 
\begin{equation} \label{eq:EeJap}
\mathcal{E}(\Sigma)\ge{2\over3}{J^2\over{2I_{s}}}
\end{equation}
where $I_s$ is the moment of inertia of a thin shell. Thus, in the linear theory, the quasi-local energy of a thin spherical shell $\mathcal{E}(\Sigma)$ is bounded from below by $2/3$ of its minimum rotational energy. This result suggest that the inequality \eqref{eq:BekJ} could be strengthened. However, the fact that the complete inequality holds for the Kerr-Newman black hole is a strong constraint to any attempt to modify it. In addition, Dain has proved \citep{Dain2015} that the inequality between energy and angular momentum is a direct consequence of the Dominant Energy Condition (DEC), and, moreover, that the equality in (\ref{eq:BekJ}) is reached in Maxwell electrodynamics only for radiation fields, i.e. $E_\alpha E^{\alpha}=B_\alpha B^{\alpha}=B_{\alpha}E^{\alpha}=0$.

There are many examples of Nonlinear Electrodynamics (NLED) in the literature~\citep{Gaete2014, Gaete2014a, Hendi2012, Kruglov2015, Kruglov2017, Dunne2004}. Up to now, Maxwell electrodynamics has never been seriously challenged by any experiment. Nevertheless, there are interesting theoretical arguments \citep{Born1933, Born410, Fradkin1985, Metsaev1987, Tseytlin1997, Gibbons2001, Abalos2015} that prompt us to investigate NLED. In addition, NLED naturally appears as the effective action for quantum electrodynamics if we consider vacuum polarization effects \citep{Euler1936, Delphenich2003}. 

Bekenstein bounds and inequalities are supposed to have universal validity. Therefore, it is reasonable to use these inequalities as a possible test for NLED candidates. This criterion can be understood as complementary to already known theoretical \citep{Goulart2009, Shabad2011, Deser1980} and experimental \citep{Ferraro2007, Ferraro2010, Flood2012, Fouche2016} criteria in the literature. The minimum requirement for a NLED is to recover Maxwell electrodynamics in the appropriate regime. However, there are physical arguments based on causality that restrict the form of NLED lagrangians. In this paper we shall use inequalities \eqref{eq:BekC}, \eqref{eq:BekQ} and \eqref{eq:BekJ} as a physical argument to test NLED. 

The paper is organised as follows. In the next section, in order to fix notation, we briefly review the covariant formalism of linear and nonlinear electrodynamics. In section \ref{BekIneq:BIF} we explicitly show that Born-Infeld Electrodynamics, similarly to Maxwell electrodynamics, also satisfy all three inequalities. In addition, in section \ref{BekIneq:VBI} we present counter-examples showing that NLED in general does not satisfy Bekenstein inequalities. In section \ref{BekIneq:CAMI} we investigate the relation of the angular momentum inequality \eqref{eq:BekJ} with causality and show that, even though  being a consequence of the DEC, this inequality cannot be strictly associated with causality. Finally in section \ref{conclusion} we conclude with some general remarks.

\section{Electrodynamics}\label{Electrod}

In this short review we shall define some relevant objects and fix our notation. Throughout our development, we shall use Heaviside-Lorentz units with $\kappa_{B}=\hbar=c=1$. Let us start by fixing spacetime as the flat Minkowski metric that in Cartesian coordinates reads $\eta_{\mu\nu}=\text{diag}\left(1,-1,-1,-1\right)$. 

Electromagnetism is understood as the vector gauge theory with symmetry group $U(1)$, and hence, described by the Faraday tensor $F_{\mu\nu}=\partial_{\mu}A_{\nu}-\partial_{\nu}A_{\mu}$. This automatically guarantees, for any electromagnetic theory, the validity of the second pair of Maxwell's equations given by $ \partial_{[\alpha}F_{\mu\nu]}=0$ where brackets means totally antisymmetry in the indices.

The electric and magnetic fields are defined as the projection of the Faraday tensor and its dual along the observer's worldline. The dual of the Faraday tensor is given by $\widetilde{F}^{\mu\nu}={1\over2}\eta^{\mu\nu\alpha\beta}F_{\alpha\beta}$ where $\eta^{\mu\nu\alpha\beta}$ is the totally antisymmetric Levi-Civita tensor. Thus, consider an observer with normalised velocity $v^\mu$, i.e. $v^{\mu}v_{\mu}=1$. The electric and magnetic fields are defined respectively as
\begin{align}\label{eq:EeB}
E^\mu=F^{\mu}_{\ \nu}v^{\nu} \quad , \qquad
B^{\mu}=\widetilde{F}^{\mu}_{\ \nu}v^{\nu} \quad .
\end{align}

Both electromagnetic vectors are spacelike with negative norms, i.e. $E^\mu E_\mu=-E^2$ and $B^\mu B_\mu=-B^2$. Furthermore, by definition they are perpendicular to the velocity field $E^\mu v_\mu=B^\mu v_\mu=0$. We can construct two Lorentz invariant quantities with the Faraday tensor and its dual, namely,
\begin{align}\label{eq:EeB}
&F\equiv{1\over2}F^{\mu\nu}F_{\mu\nu}=E_{\alpha}E^{\alpha}-B_{\alpha}B^\alpha \\
&G\equiv {1\over4}\widetilde{F}^{\mu\nu}F_{\mu\nu}=B_{\alpha}E^{\alpha}
\end{align}

These Lorentz invariants constitute the only linearly independent scalars that can be constructed from $F^{\mu\nu}$ and its dual \citep{Landau1980}. Indeed, a direct calculation shows the following algebraic relations
\begin{eqnarray}
&&\widetilde{F}^{\mu \alpha}\widetilde{F}_{\alpha \nu}-F^{\mu \alpha}F_{\alpha \nu}=F \delta^\mu{}_{\nu}\\
&&\widetilde{F}^{\mu \alpha}F_{\alpha \nu}=-G \delta^\mu{}_{\nu}\\
&&F^\mu{}_\alpha F^\alpha{}_\beta F^\beta{}_\nu=-G \widetilde{F}^\mu{}_{\nu}-F\ F^\mu{}_{\nu}\\
&&F^\mu{}_\alpha F^\alpha{}_\beta F^\beta{}_\lambda F^\lambda{}_\nu=G^2\ \delta^\mu{}_{\nu}-F\ F^\mu{}_{\alpha}F^\alpha{}_{\nu} \quad .
\end{eqnarray}
Therefore, one can construct rank-2 objects only up to second power of the Faraday tensor, i.e. any power of the electromagnetic tensor and its dual is a combination of the identity $\delta^\mu{}_{\nu}\, , \, F^\mu{}_{\nu}$,  $\widetilde{F}^\mu{}_{\nu}$ and $ F^\mu{}_{\alpha}F^\alpha{}_{\nu}$.

The source of electrodynamics is charged particles. We shall denote $\Sigma$ as the region which contains all charges. 
\begin{df}\label{defR}
The size of the region $\Sigma$ can be characterised by the radius $\mathcal{R}$, which we define as the radius of the smallest sphere $\mathcal{B_R}$ that encloses $\Sigma$. Additionally, we shall designate the center of this sphere by $x_0$.
\end{df}

Thus, the total electric charge contained in $\Sigma$
\[
Q(\Sigma)=\int_\Sigma{\rho}\quad .
\]

Two other important quantities for our analysis are the energy and angular momentum of the distribution of charges. These quantities are defined as integral of combinations of the energy-momentum tensor components. We shall define our energy-momentum tensor through the variation of the matter action with respect to the metric tensor. Thus, we have
\begin{equation}\label{EMT:generic}
T_{\mu\nu}=\frac{2}{\sqrt{-g}}\frac{\delta}{\delta g^{\mu\nu}}(\sqrt{-g}\mathcal{L}_{\rm mat})=2\frac{\delta \mathcal{L}_{\rm mat}}{\delta g^{\mu\nu}}-\mathcal{L}_{\rm mat} \, g_{\mu\nu}\quad .
\end{equation}

\subsection{Maxwell Electrodynamics}\label{Intro:MED}

Maxwell's electrodynamics is described by a set of four differential equations. The two source-free equations, allow us to define the Faraday tensor as the exterior derivative of the vector potential 1-form, i.e. $F={\rm d} A$. The other two equations are associated with the source terms.  Defining the current vector $j^\mu=(\rho, \mathbf{j})$, the second set of Maxwell's equations reads
\begin{equation}
\partial_\mu F^{ \mu \nu}=j^\nu \label{max1} \quad .\\
\end{equation}

These equations can be derived from a variational principle for the vector potential $A_\mu$ where the appropriate action is defined with the Larmor lagrangian, i.e. $\mathcal{L}=-{1\over2}F$. The invariant $G$ represents a total divergence and hence does not contribute to the dynamics equations. Thus, up to a multiplicative constant, the Larmor lagrangian is the unique linear electromagnetic lagrangian.

As usual, the energy-momentum tensor is defined as the variation of the matter action with respect to the spacetime metric, which for Maxwell's theory gives
\begin{equation} \label{eq:TM}
T^\mu{}_{\nu}=F^{\mu \alpha}F_{\alpha \nu}+{F\over2}\delta^\mu{}_{\nu} \quad ,
\end{equation}

In particular, the maxwellian electromagnetic energy density $u_M$ is the time-time components of the energy-momentum tensor, i.e. $u_{M}=-\frac12  \left(E_{\alpha}E^{\alpha}+B_{\alpha}B^{\alpha}\right)$ and the total energy reads
\begin{equation} \label{eq:EM}
\mathcal{E}_M=-{1\over2}\int_{\mathbb{R}^3}\left(E_{\alpha}E^{\alpha}+B_{\alpha}B^{\alpha}\right)\ .
\end{equation}

Similarly, the angular momentum of a region $\Sigma$ with respect to a point $x_0$ projected along the direction $\mathbf{k}$ is defined as
\begin{equation} \label{eq:JM}
J(\Sigma)=\int_{\Sigma}\epsilon_{ijk}\epsilon^{iab}E_{a}B_{b}k^{j}x^{k}\ .
\end{equation}
\subsection{Nonlinear Electrodynamics}\label{Intro:NLED}

The most general Lorentz invariant electromagnetic lagrangian is a function of the two scalar invariants $F$ and $G$, i.e. $\mathcal{L}=\mathcal{L}(F,G)$.  Given an arbitrary lagrangian, its energy-momentum tensor reads
\begin{equation} \label{eq:TLFG}
T^\mu{}_{\nu}=-F^{\mu\alpha}E_{\alpha \nu}-\mathcal{L}\, \delta^\mu{}_{\nu}
\end{equation}
where the excitation tensor is defined as $E_{\mu\nu}\equiv {\partial\mathcal{L}\over{\partial F^{\mu\nu}}}=2\left(\mathcal{L}_{F}F_{\mu\nu}+\mathcal{L}_G\widetilde{F}_{\mu\nu}\right)$, and $\mathcal{L}_{x}$ stands for partial derivative of the lagrangian with respect to $x$. The field equations can be written in terms of $E_{\mu\nu}$ as
\begin{align}\label{eq:fieldLFG}
\partial_{\mu}E^{\mu\nu}=-J^{\nu}
\end{align}

In the same way as before, the energy density reads
\begin{equation} \label{eq:uLFG}
u=2\left(\mathcal{L}_{F}E^\alpha E_\alpha+\mathcal{L}_G G\right)-\mathcal{L}\  ,
\end{equation}
and the angular momentum is
\begin{equation} \label{eq:MAFG}
J(\Sigma)=-2\int_{\Sigma}\mathcal{L}_{F}\epsilon_{ijk}\epsilon^{iab}E_aB_bk^{j}x^{k}\ .
\end{equation}

It is worth noting that, contrary to the energy density, the angular momentum depends only on the first derivative of the lagrangian with respect to the invariant $F$. Evidently, both expressions recover the linear case for $\mathcal{L}=-\frac12 F$.

\section{Bekenstein Bounds and Inequalities Within NLED}\label{BekIneq}

There are two possible ways to approach the interplay of NLED and Bekenstein bounds and inequalities. From one point of view, since the latter is supposed to be valid for an arbitrary physical system, they can be used to test possible NLED candidates. On the other hand, NLED is a fertile framework that allows us to investigate different theoretical situations, which can provide us with deeper insights on the physical meaning of the Bekenstein inequalities.

In this section, we begin by proving that Born-Infeld electrodynamics satisfies all three inequalities. Our development follows closely the analysis done by Dain in \cite{Dain2015} for Maxwell's electrodynamics. Next we show a concrete example of NLED that violates the inequality between energy and charge but still respects the inequality between energy and angular momentum.

\subsection{Born-Infeld Electrodynamics}\label{BekIneq:BIF}

Maxwell's electrodynamics is a classical theory that suffers from divergences such as the value of the electromagnetic energy as one approaches a charged particle. This kind of problem motivated finite size models for the electron, which would give an upper bound for its self-energy as one probes the radius goes to zero limit. An alternative context is to modify the electrodynamics to include nonlinear effects. The first attempt along these lines, due to G. Mie \cite{Mie1912}, was to introduce a model where there is an upper limit for the value of the electric field but this formulation was not Lorentz covariant.

Following Mie, in 1933, M. Born and L. Infeld proposed a nonlinear modification of Maxwell electrodynamics \citep{Born1933, Born410, Born425} that also has an upper limit for the electromagnetic fields. Born-Infeld electrodynamics is a special nonlinear theory due to its theoretical features. By construction, it is a gauge invariant theory with finite electromagnetic mass point-like sources. The energy is positive definite and the Poynting vector is everywhere non-spacelike. In addition it has no birefringence phenomena. 

In general, photon propagation in nonlinear theories depends on the value of the electromagnetic fields. As a consequence, different polarisation states propagate along different light-cones~\cite{Plebanski1970, boillat1966, boillat1967, DeLorenci2000, Chernitskii1998}. Notwithstanding, G. Boillat~\cite{boillat1970, boillat1972} showed that Born-Infeld is unique in the sense that is the only NLED without birefringence phenomena and shock waves can occur only across characteristic surfaces of the field equations as is the case for the linear theory. 

Besides trying to eliminate the classical divergences, Born and Infeld proposal was inspired by the theory of general relativity. They argued that the diffeomorphism invariance of the action can be obtained by taking the square root of the determinant of a tensor field $|a_{\mu \nu}|$. In particular, they identified its symmetric part with the metric tensor and its antisymmetric part with the Faraday tensor, i.e. $a_{\mu \nu}=g_{\mu \nu}+F_{\mu \nu}$. In order to recover Maxwell electrodynamics in the weak field limit, the desired combination is
\begin{equation} \label{eq:LBIGR}
S=\int{}{\rm d}\tau \ \beta^2 \left(\sqrt{-|g_{\mu \nu}|}-\sqrt{-|g_{\mu \nu}+\beta^{-1}F_{\mu \nu}|}\ \right)\quad ,
\end{equation}
where $\beta$ constitutes a maximum field parameter. Assuming Cartesian coordinates in a flat spacetime, the above action reads
\begin{equation} \label{eq:LBI}
S=\int{}{\rm d}\tau \ \beta^2\left(1-\sqrt{U}\right)
\end{equation}
where $U=1+F/\beta^2-G^2/\beta^4$.

\begin{figure}[ht]
\centering
\includegraphics[width=0.45\textwidth]{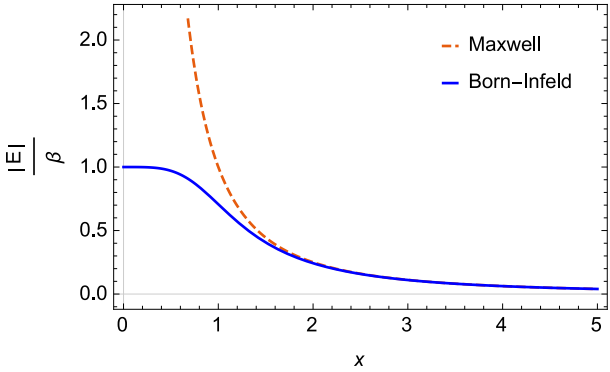}
\caption{The electrostatic field of a pointlike charged particle as a function of $x=r\sqrt{\beta/e}$ for Born-Infeld (solid line) and Maxwell (dashed line) electromagnetism.}
\label{fig:EBI}
\end{figure}

The Born-Infeld field equations read
\begin{equation} \label{eq:fieldBI}
\partial_{\mu}\left[{1\over\sqrt{U}}\left(-F^{\mu\nu}+{G\over{\beta^2}}\widetilde{F}^{\mu\nu} \right) \right]=-j^{\nu}\ ,
\end{equation}
which constitute the generalisation for the Amp\`{e}re-Maxwell and Gauss' equations. These equations can be recast in vector notation as
\begin{subequations} \label{eq:fieldBI1}
\begin{eqnarray}
\nabla\cdot\mathbf{D}=\rho \label{eq:GaussBI} \\
\nabla\times\mathbf{H}-{\partial\mathbf{D}\over{\partial{t}}}=\mathbf{j} \label{eq:AMBI}
\end{eqnarray}
\end{subequations}
with 
\begin{align}
\mathbf{D}\equiv {1\over{\sqrt{U}}}\left(\mathbf{E}+{\left(\mathbf{E}\cdot\mathbf{B}\right)\over{\beta^2}}\mathbf{B} \right)
\ , \ \mathbf{H}\equiv {1\over{\sqrt{U}}}\left(\mathbf{B}-{\left(\mathbf{E}\cdot\mathbf{B}\right)\over{\beta^2}}\mathbf{B} \right) \ , 
\end{align}
which resemble Maxwell's equations inside matter with nonlinear permittivity and permeability. For the electrostatic case, eq.~\eqref{eq:GaussBI} allow us to calculate the electric field for a pointlike charged particle. The value of the electric field in the limit $r\rightarrow 0$ gives the maximum electrostatic field $\beta$. Figure~\ref{fig:EBI} shows the difference between the Born-Infeld and Maxwell electrostatic field.

Using the Born-Infeld lagrangian eq.~\eqref{eq:LBI} in the definition eq.~\eqref{eq:TLFG}, we obtain the energy-momentum tensor
\begin{equation} \label{eq:TBI}
T_{\mu\nu}={1\over\sqrt{U}}\left(F_{\mu\alpha}F^{\alpha}_{\ \nu}+{G^2\over{\beta^2}}g_{\mu\nu}\right)+g_{\mu\nu}\beta^2\left(\sqrt{U}-1\right)\ .
\end{equation}

A similar calculation gives the angular momentum of the distribution of charged particles in the region $\Sigma$ as
\begin{equation} \label{eq:JBI}
J_{BI}(\Sigma)=\int_{\Sigma}{1\over\sqrt{U}}\epsilon_{ijk}\epsilon^{iab}E_{a}B_{b}k^{j}x^{k}\ ,
\end{equation}
and its energy density
\begin{equation} \label{eq:uBI}
u_{BI}={\beta^2 \over{\sqrt{U}}}\left(1-\sqrt{U}-\frac{B_{\alpha}B^{\alpha}}{\beta^2}\right) \ .
\end{equation}

Born-Infeld electrodynamics has a maximum value for both fields given by the parameter $\beta$. Thus, for future analysis, it is convenient to normalise the electric and magnetic field, i.e. we define the two parameters $\alpha\equiv \beta^{-1}|\mathbf{E}|$ and $\gamma\equiv \beta^{-1}|\mathbf{B}|$ that give, respectively, the normalised strength of the electric and magnetic fields and such that $(\alpha, \gamma) \in [0,1]$. In terms of these parameters the Born-Infeld function reads
\begin{equation}\label{defU}
U=1+\gamma^2-\alpha^2-\gamma^2\alpha^2\cos^2 \theta\ ,
\end{equation}
with $\cos \theta \equiv \mathbf{E}.\mathbf{B}/\left(|\mathbf{E}||\mathbf{B}|\right)$. Note that $U \in [0,2]$ but can be divided in two distinct domains. As a fact, $U \in [0,1)$ implies $\alpha> \gamma$ and $U \in [1,2]$ implies $\gamma\geq \alpha$.

\subsubsection{Inequality Between Charge and Energy}
We will begin by examining the inequality between charge and energy eq.~\eqref{eq:BekQ} in Born-Infeld electrodynamics. We shall prove the following theorem.
\begin{thm}\label{thmQE}
Assume that the charge density $\rho$ has compact support contained in the region $\Sigma$ and Born-Infeld electrodynamics holds. Then, the total charge $Q$ contained in $\Sigma$ and the total electromagnetic energy $\mathcal{E}_{BI}$ of the system satisfy the inequality
\begin{equation}\label{eqthmQE}
\mathcal{E}_{BI}> \frac{Q^2}{8\pi\mathcal{R}}
\end{equation}
where $\mathcal{R}$ is defined as in definition \ref{defR}.
\end{thm}

\textit{Proof.}---In theorem 2.2 of ~\cite{Dain2015}, Dain has shown\footnote{There is a factor $4\pi$ of difference due to our choice of units. Namely, eq.~(4) in \cite{Dain2015} has a factor $4\pi$ that does not appear in our eq.~\eqref{eq:fieldBI1}.} that Maxwell's electrodynamics satisfies a similar inequality, namely, that
\begin{equation}\label{eqthmQEMax}
\mathcal{E}_{Ms}\geq \frac{Q^2}{8\pi\mathcal{R}}
\end{equation}
where $\mathcal{E}_{Ms}$ is the Maxwell electrostatic energy of the system and $Q$ and $\mathcal{R}$ have the same meaning as here. Furthermore, there is a rigidity condition. Equality in \eqref{eqthmQEMax} holds if and only if the electric field is the one produced by a spherical thin shell of constant surface charge density and radius $\mathcal{R}$. As a consequence, for the equality to hold in \eqref{eqthmQEMax}, the electric field has to vanish inside $\Sigma$.

In order to prove ineq.~\eqref{eqthmQE} it is sufficient to show that the Born-Infeld energy is always greater than its electrostatic counterpart and then show that the Born-Infeld electrostatic energy is always greater than the Maxwell electrostatic energy.

In Maxwell's linear theory, the electromagnetic energy is always greater than or equal to the electrostatic case but this is no longer the case for a generic NLED. The Born-Infeld theory is a special case where this property is indeed valid. Note that, since $1-\alpha^2\cos^2 \theta \geq 0$, Born-Infeld energy density is an increasing function of the parameter $\gamma$ and hence
\begin{equation}\label{proofQESpt1}
u_{BI}(\alpha,\gamma)\geq u_{BI}(\alpha,0)\quad .
\end{equation}

As a consequence the Born-Infeld energy is always greater than its electrostatic version. Thus, it suffices to show that Born-Infeld electrostatic energy density is always greater than Maxwell electrostatic energy density. Their difference reads
\begin{align}\label{proofQESpt2}
u_{BI}(\alpha,0)-u_{Ms}&=\frac{\beta^2}{\sqrt{U}}\left[1-\left(1+\frac{\alpha^2}{2}\right)\sqrt{1-\alpha^2}\right]\\
&\geq \frac{\beta^2}{\sqrt{U}}\left[1-\left(1+\alpha^2\right)\sqrt{1-\alpha^2}\right]\\
&=\frac{\beta^2}{\sqrt{U}}\left[1-\sqrt{\frac{1-\alpha^4}{1+\alpha^2}}\right]\geq 0
\end{align}

The equality above holds only when the electric field vanishes everywhere. Therefore, the Born-Infeld electrostatic energy is always greater than the Maxwell electrostatic energy. \hfill$\square$\\

There is no rigidity statement for Born-Infled electrodynamics because its energy density is always greater than Maxwell. As we have mentioned before, in NLED, the nonlinearity of the theory allows for a non-trivial dependence of its energy density with the strength of the electromagnetic fields. Thus, it is possible to have NLED with energy density lower than the Maxwell energy density. We will explore this scenario later on.

\subsubsection{Inequality Between Energy and Angular Momentum} \label{sec:EJBI}

Our next step is to prove the inequality between energy and angular momentum. The main difference to theorem \ref{thmQE} is that inequality \eqref{eq:BekJ} relates two quasi-local quantities. In addition, in this section we shall consider the case $Q=0$ and $J\neq0$ but otherwise arbitrary electromagnetic field's configurations. We want to prove the following theorem.

\begin{thm}\label{thmJE}
Consider a distribution of charged particles in the region $\sigma$ with no net charge, i.e. $Q=0$. Let the radius $\mathcal{R}$ to be defined as in \ref{defR} and $x_0$ the center of the corresponding sphere. If Born-Infeld electrodynamics equations hold, then 
\begin{equation}\label{eqthmJE}
\mathcal{E}_{BI}(\Sigma)\, \mathcal{R} \ge \vert J_{BI}(\Sigma)\vert \ ,
\end{equation}
where $J_{BI}(\Sigma)$ is the angular momentum of the electromagnetic field given by eq.~\eqref{eq:JBI} with respect to the point $x_0$. Furthermore, the equality in eq.~\eqref{eqthmJE} holds if and only if the electromagnetic fields vanish in $\Sigma$.
\end{thm}

\textit{Proof.}---In order to prove the above theorem we shall calculate the difference between the energy and angular momentum in the region $\Sigma$. Using the definitions eq.~\eqref{eq:JBI} and \eqref{eq:uBI}, we have 
\begin{align} \label{eq:EMABI1}
\mathcal{E}_{BI}(\Sigma)-{1\over\mathcal{R}}\vert J_{BI}(\Sigma)\vert=&\int_{\Sigma}{\beta^2\over{\sqrt{U}}}\left(1+\gamma^2-\sqrt{U}\right)+\nonumber\\
&-{1\over\mathcal{R}}\left|\int_{\Sigma}\left({1\over\sqrt{U}}\epsilon_{ijk}\epsilon^{iab}E_{a}B_{b}k^{j}x^{k}\right) \right|\nonumber\\
&\ge\int_{\Sigma}{\beta^2\over\sqrt{U}}\left( 1+\gamma^2 -\sqrt{U} -{x\over{\mathcal{R}}}\alpha\gamma \right)\ .
\end{align}
In the last line we have used the inequality $\left| \int f(x)\right| \le\int\vert f(x)\vert$ and the fact that $\Big|\big(\mathbf{x} \times (\mathbf{E} \times \mathbf{B})\big) \cdot \hat{\mathbf{k}}\Big|\leq\Big|\mathbf{x} \times (\mathbf{E} \times \mathbf{B})\Big|\leq |\mathbf{x}||\mathbf{E}||\mathbf{B}|$ .
Recalling eq.~\eqref{defU} we can rearrange the above expression as 
\begin{align} \label{eq:EMAbi5}
\mathcal{E}_{BI}(\Sigma)-{1\over\mathcal{R}}\vert J_{BI}(\Sigma)\vert\geq&\int_{\Sigma}{\beta^2\over{2\sqrt{U}}}\Bigg[\left(1-\sqrt{U}\right)^2+\left(\gamma-{x\over{\mathcal{R}}}\alpha\right)^2+
\nonumber\\
&+\alpha^2\gamma^2\cos^2\theta+\alpha^2\left(1-{x^2\over{\mathcal{R}^2}}\right) \Bigg]
\end{align}
It is obvious that all the integrands in the above equation are nonnegative and hence the integral is greater or equal to zero. Thus, we have proved inequality \eqref{eqthmJE}. In the above form, it can also be seen that the equality can only be achieved when the integrand in \eqref{eq:EMAbi5} is zero, hence, every term in the integrand has to identically vanish. Thus, equality holds if and only if the electric and magnetic fields vanish in $\Sigma$ proving the rigidity condition. \hfill$\square$\\

\subsubsection{Inequality Between Charge, Energy and Angular Momentum}

Finally we shall prove the full inequality \eqref{eq:BekC} involving charge, angular momentum and total energy of the system.

\begin{thm}\label{thmQJE}
Assume that the charge density $\rho(x,t_0)$, for some time $t_0$, has compact support contained in the region $\Sigma$. Consider a solution of Born-Infeld dynamics equations that decays at infinity. Then, at $t_0$, the total charge $Q$ contained in $\Sigma$, the total electromagnetic energy $\mathcal{E}_{BI}$ and the angular momentum $J_{BI}(\Sigma)$ with respect to $x_0$ satisfy the inequality
\begin{equation}\label{eqthmQJE}
\mathcal{E}_{BI}>\frac{Q^2}{8\pi\mathcal{R}}+\frac{\left|J_{BI}(\Sigma)\right|}{\mathcal{R}} \ ,
\end{equation}
where $\mathcal{R}$ and $x_0$ are defined as in definition \ref{defR}.
\end{thm}

\textit{Proof.}--- Let us express the electric and magnetic fields in the Coulomb gauge 
\begin{align}\label{Cgauge}
&\mathbf{B}=\nabla\times\mathbf{A}\ , &\mathbf{E}=-\mathbf{\nabla}\Phi-\partial_{t}\mathbf{A}\ ,
\end{align}
where the vector potential satisfies the Coulomb gauge condition
\begin{align}\label{Cgauge2}
\nabla\cdot\mathbf{A}=0
\end{align}
It is convenient to decompose the scalar potential using an auxiliary potential $\Phi_1$. Thus, we define
\begin{equation} \label{eq:pot}
\Phi=\Phi_0+\Phi_1
\end{equation}
where
\begin{equation} \label{eq:potes}
\Phi_0=
\begin{cases}
Q/ {4\pi r}, &\text{if $r\ge \mathcal{R}$}\ ,\\
Q/ 4\pi\mathcal{R}, &\text{if $r\le\mathcal{R}$} \ ,\\
\end{cases}
\end{equation}
is the potential of a spherical shell of radius $\mathcal{R}$ and the same total charge $Q$ as contained in $\Sigma$. Note that $\nabla\Phi_0=0$ inside $\mathcal{B_R}$ and by construction we have 
\begin{align}\label{propphi1}
\Delta \Phi_1=
\left\{\begin{array}{cl}
0 & \mbox{if}\ r> \mathcal{R}\\
-\rho & \mbox{if}\ r< \mathcal{R}\\
\end{array}
\right.
\end{align}
and
\begin{align}\label{prophi12}
\oint_{\partial \mathcal{B_R}} \partial_r \Phi_1=0\ .
\end{align}

Before calculating the total energy $\mathcal{E}_{BI}$, let us consider the integral over all space of the modulus squared of the electric field.
\begin{equation}\label{intE2}
\int_{\mathbb{R}^3}\mathbf{E}^2=\int_{\mathbb{R}^3}\left\{\vert\nabla\Phi\vert^2+\vert\partial_{t}\mathbf{A}\vert^2+2\nabla\Phi\cdot\partial_{t}\mathbf{A}\right\} \ .
\end{equation}

The last term gives no contribution since it can be recast as a surface term. Indeed, we can rewrite it as 
\begin{equation}\label{intphiA}
\int_{\mathbb{R}^3}\nabla\Phi\cdot\partial_{t}\mathbf{A}=
\int_{\mathbb{R}^3}\Big[ \nabla \cdot \left(\Phi \partial_{t}\mathbf{A}\right)-\Phi \partial_{t} \left(\nabla \cdot\mathbf{A}\right)\Big]=0
\end{equation}
where the first term on the right is zero due to Gauss' theorem and the falloff condition of $\Phi$, and the last term vanishes since the vector potential satisfies the Coulomb gauge condition.

Now we shall use the auxiliary scalar potential to rewrite the first term of eq.~\eqref{intE2}.
\begin{equation}\label{intphi2}
\int_{\mathbb{R}^3}\vert\nabla\Phi\vert^2=\int_{\mathbb{R}^3}\vert\nabla\Phi_0\vert^2+\vert\nabla\Phi_1\vert^2+2\nabla\Phi_0\cdot \nabla\Phi_1
\end{equation}
Again, the last term does not contribute. We can decompose the integral in two regions: inside and outside of the sphere $\mathcal{B_R}$. Inside the sphere
\begin{equation}
\int_{\mathcal{B_R}}\nabla\Phi_0\cdot \nabla\Phi_1=0
\end{equation}
since the potential $\Phi_0$ is constant and hence $\nabla\Phi_0=0$. Outside the sphere we have
\begin{align}
\int_{\mathbb{R}^3\backslash \mathcal{B_R}}{\nabla\Phi_0\cdot \nabla\Phi_1}&=
\int_{\mathbb{R}^3\backslash \mathcal{B_R}}
\Big[\nabla\cdot\left( \Phi_0\nabla\Phi_1\right)-\Phi_0\Delta\Phi_1\Big]\nonumber\\
&=\int_{\mathbb{R}^3\backslash \mathcal{B_R}}
\nabla\cdot\left( \Phi_0\nabla\Phi_1\right)\ ,
\end{align}
where we have used eq.~\eqref{propphi1}. Gauss' theorem now gives
\begin{align}
\int_{\mathbb{R}^3\backslash \mathcal{B_R}}
\nabla\cdot\left( \Phi_0\nabla\Phi_1\right)&=\lim_{r\rightarrow \infty}\oint_{\partial\mathcal{B}_r}\Phi_0\partial_r\Phi_1-\oint_{\partial\mathcal{B_R}}\Phi_0\partial_r\Phi_1\nonumber\\
&=-\oint_{\partial\mathcal{B_R}}\Phi_0\partial_r\Phi_1\nonumber\\
&=-\Phi_0\oint_{\partial\mathcal{B_R}}\partial_r\Phi_1=0\ .
\end{align}
From the first to the second line we have used the falloff condition of the potential. From the second to the third line we have used the fact that $\Phi_0$ is constant on the sphere $\mathcal{B_R}$, and finally in the last line eq.~\eqref{prophi12} shows that the cross term does not contribute to eq.~\eqref{intphi2}.

Our last step is to combine $\vert\nabla\Phi_1\vert^2$ with $\vert\partial_{t}\mathbf{A}\vert^2$
\begin{align}
&\int_{\mathbb{R}^3}\vert\nabla\Phi_1\vert^2+\vert\partial_{t}\mathbf{A}\vert^2=\int_{\mathbb{R}^3}\vert\nabla\Phi_1+\partial_{t}\mathbf{A}\vert^2-2\nabla\Phi_1\cdot\partial_{t}\mathbf{A}\nonumber\\
&=\int_{\mathbb{R}^3}\vert\nabla\Phi_1+\partial_{t}\mathbf{A}\vert^2\ ,
\end{align}
where we have discarded the cross term with the same arguments as used in eq.~\eqref{intphiA}. Combining all these results, the integral of the modulus squared of the electric fields yields 
\begin{align}\label{intE22}
\int_{\mathbb{R}^3}\mathbf{E}^2=&\int_{\mathbb{R}^3}\vert\nabla\Phi_0\vert^2+\vert\nabla\Phi_1+\partial_{t}\mathbf{A}\vert^2\nonumber\\
=&\frac{Q^2}{4\pi\mathcal{R}}+
\int_{\mathbb{R}^3}\vert\nabla\Phi_1+\partial_{t}\mathbf{A}\vert^2\ .
\end{align}

The Born-Infled total energy reads
\begin{equation}\label{pfEQJ1}
\mathcal{E}_{BI}=\int_{\mathbb{R}^3}{\beta^2 \over{\sqrt{U}}}\left(1+\gamma^2-\sqrt{U}\right)\ .
\end{equation}

We can sum and subtract the integral eq.~\eqref{intE2} to obtain
\begin{align}\label{pfEQJ2}
\mathcal{E}_{BI}
=&\int_{\mathbb{R}^3}\frac{\mathbf{E}^2}{2}+\int_{\mathbb{R}^3}{\beta^2 \over{\sqrt{U}}}\left(1+\gamma^2-\sqrt{U}\left(1+\frac{\alpha^2}{2}\right)\right)\nonumber\\
=&
\frac{Q^2}{8\pi\mathcal{R}}+
\int_{\mathbb{R}^3}{\beta^2 \over{\sqrt{U}}}\left(1+\gamma^2-\sqrt{U}\left(1+\frac{\alpha^2}{2}\right)\right)\nonumber\\
&+\frac12\int_{\mathbb{R}^3}\vert\nabla\Phi_1+\partial_{t}\mathbf{A}\vert^2
\end{align}

Now, we can split the limits of integration to separate the space in inside and outside $\Sigma$. Thus, we have
\begin{align}\label{pfEQJ3}
\mathcal{E}_{BI}
=&
\frac{Q^2}{8\pi\mathcal{R}}+\mathcal{E}_{BI}(\Sigma)+\int_{\mathbb{R}^3\backslash \Sigma}\Bigg[f(\alpha,\gamma)+\frac{\vert\nabla\Phi_1+\partial_{t}\mathbf{A}\vert^2}{2}\Bigg]
\end{align}
where
\begin{equation}\label{funcf1}
f(\alpha,\gamma)\equiv {\beta^2 \over{\sqrt{U}}}\left(1+\gamma^2-\sqrt{U}\left(1+\frac{\alpha^2}{2}\right)\right)\ .
\end{equation}

Theorem \ref{thmJE} allows us to write
\begin{align}\label{pfEQJ4}
\mathcal{E}_{BI}-\frac{Q^2}{8\pi\mathcal{R}}-\frac{\vert J(\Sigma)\vert}{\mathcal{R}}\geq
\int_{\mathbb{R}^3\backslash \Sigma}\Bigg[f(\alpha,\gamma)+\frac{\vert\nabla\Phi_1+\partial_{t}\mathbf{A}\vert^2}{2}\Bigg]\ .
\end{align}

Note that the function $f(\alpha,\gamma)$ is nonnegative. Rearranging the terms we have
\begin{align}\label{argfuncf1}
1+\gamma^2-\sqrt{U}\left(1+\frac{\alpha^2}{2}\right)=&\frac12\left[
\left(1-\sqrt{U}\right)^2+\alpha^2\gamma^2\cos^2\theta \right.\nonumber\\
&\left.+\alpha^2\left(1-\sqrt{U}+\frac{\gamma^2}{\alpha^2}\right)
\right]
\end{align}
The only term that is not explicitly nonnegative on the righthand side of the above equation is the last one. This comes from the fact that $0\leq U\leq 2$. However, one can check that $U>1$ only if $\gamma>\alpha$. Thus, we have $1-\sqrt{U}+\gamma^2/\alpha^2>0$, which in turn implies $f(\alpha,\gamma)>0$. Therefore, the integrand on righthand side of inequality~\eqref{pfEQJ4} is nonnegative. Furthermore, the integral has to be greater than zero since in order to have $\nabla\Phi_1=0$ we need a nonzero electric. Indeed, either $\alpha=0$ and $\nabla\Phi_1=-\nabla\Phi_0$ or $\nabla\Phi_1=0$ and $\alpha\neq0$.  \hfill$\square$\\

Born-Infeld electrodynamics is an important example of NLED. This theory, besides respecting the Maxwell limit, has many unique features. We have proofed that Born-Infeld electrodynamics satisfies all three Bekenstein inequalities. Assuming the universal validity of these inequalities, Born-Infeld can include this extra feature to its theoretical motivations.

\subsection{Breaking Bekenstein's Inequality}\label{BekIneq:VBI}

In this section we want to show a specific NLED that violates the simplest of the three inequalities, namely the one relating the total energy and total charge of the system. 

In Maxwell's theory, this inequality can be interpreted as showing that the total energy is always greater than the minimum electrostatic energy given by a thin spherical shell charge distribution. Thus, it might seem that any NLED would also satisfy this inequality. Indeed, many NLED do satisfy it~\citep{Born425, Hendi2012, Gaete2014}. In order to violate inequality (\ref{eq:BekQ}) we need a NLED that has a minimum electrostatic energy lower than the minimum maxwellian electrostatic energy.

Thus, consider the NLED $\mathcal{L}(F)$ given by the logarithmic function
\begin{equation}
\mathcal{L}=\beta^2\ln\left(1-{F\over{2\beta^2}}\right)\ ,
\end{equation}
which is a modification of the logarithmic lagrangian introduce by Gaete and Helay\"el-Neto in~\citep{Gaete2014}. This NLED has the correct maxwellian limit for weak fields and hence can be considered as a physically reasonable theory. Its energy-momentum tensor reads
\begin{equation}\label{TEMLog}
T_{\mu\nu}={F_{\mu\alpha}F^{\alpha}_{\ \nu}\over{1-{F\over{2\beta^2}}}}-g_{\mu\nu}\beta^2\ln\left(1-{F\over{2\beta^2}} \right)
\end{equation}
and its electrostatic energy density is
\begin{equation}\label{uLog}
u_{\text{log}}=\beta^2\left[\frac{2\alpha^2}{2+\alpha^2}-\ln\left(1+\frac{\alpha^2}{2} \right)\right]\ .
\end{equation}

Contrary to the Born-Infeld case, in logarithmic electrodynamics there is no upper limit for the electric field, i.e. $\alpha\in[0,\infty)$. However, this NLED is well defined only together with a finite size theory of charged particles. This is due to the fact that there is a minimum allowed radius in order to guarantee the reality of the electric field. In the electrostatic regime, the electric displacement $\mathbf{D}$ is related with the electric field by
\begin{equation}\label{eqrelDE}
\mathbf{D}=\frac{\mathbf{E}}{1+\vert\mathbf{E}\vert^2/2\beta^2}\ .
\end{equation}

\begin{figure}[ht]
\centering
\includegraphics[width=0.45\textwidth]{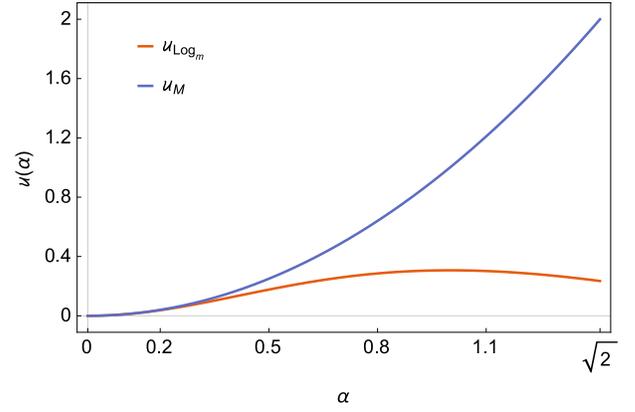}
\caption{The electrostatic energy density for logarithmic electrodynamics and the electrostatic energy density for Maxwell electrodynamics.}
\label{fig:logmod}
\end{figure}

For a static spherically symmetric distribution with total charge $Q$, Gauss' theorem shows that $\mathbf{D}=Q/4\pi r^2\, \hat{\mathbf{r}}$. We can invert eq.~\eqref{eqrelDE} and write
\begin{equation}\label{eqrelDE2}
\vert\mathbf{E}\vert=\frac{\sqrt{2}\beta}{r_0}\left(r^2\pm\sqrt{r^4-r_0^4}\right)\ ,
\end{equation}
where $r_0^2\equiv \sqrt{2}Q/4\pi \beta$ is the minimum size of charged particles. A comparison between the logarithmic and Maxwell energy density is plotted in FIG. \ref{fig:logmod}. 

The binding energy for a spherical shell of radius $r_0$ and total charge $Q$ within Maxwell electrodynamics is
\begin{equation} \label{eq:Er0M} 
\mathcal{E}_{\text{M}}={Q^2\over{8\pi r_0}}
\sim0.1186\sqrt{Q^3\beta} \ .
\end{equation}

The binding energy for the same charge distribution in the logarithmic electrodynamics can be calculated by the integral of the energy density
\begin{align}\label{eq:logmod2}
\mathcal{E}_{\text{log}}=&4\pi\int_{r_0}^{\infty}r^2dr\left\{{E^2\over{1+{E^2\over{2\beta^2}}}}-\beta^2\ln\left(1+{E^2\over{2\beta^2}} \right)\right\}\nonumber\\
=&\sqrt{\frac{Q^3\beta}{8\pi\sqrt{2}}}\int_{\sqrt{2}}^{\infty}dy y^2\left\{\frac{2}{1+y^4}-\ln\left(1+{1\over{y^4}}\right)\right\}  \nonumber\\
&\sim 0.1108\sqrt{Q^3\beta}
\end{align}

Note that the Maxwell's binding energy is greater than the logarithmic binding energy
\begin{equation}
\mathcal{E}_{\text{M}}>\mathcal{E}_{\text{log}}
\end{equation}

Since the equality in (\ref{eq:BekQ}) is reached only by $\mathcal{E}_{\text{M}}$, we can conclude that the electrostatic configuration $\mathcal{E}_{\text{log}}$ violates the inequality between energy and charge. Furthermore, following the same reasoning used in Sec.~\ref{sec:EJBI}, one can show that the logarithmic electrodynamics does satisfy the inequality between energy and angular momentum. This result proves that the validity of each partial inequality is independent of each other.

\section{Causality and the Angular-Momentum Inequality}\label{BekIneq:CAMI}

There is an interesting connection between theorem \ref{thmJE} and the spacetime causal structure. Dain has shown~\citep{Dain2015}, and we reproduce the argument in appendix \ref{app:DEC}, that the dominant energy condition (DEC) is a sufficient condition for inequality \eqref{eqthmJE}. DEC is a physically motivated condition on the energy-momentum tensor, which prohibits superluminal propagation~\citep{Wald1984, Hawking1973}.

On the other hand, inequality \eqref{eqthmJE} relates the quasi-local total energy of the system with the quasi-local angular-momentum with respect to the origin of the minimum sphere that surrounds the system. In a sense, this inequality shows that the total energy has to be greater or equal to the angular kinetic energy of the system. Furthermore, their ratio is proportional to a mean angular velocity of the system, and hence, inequality \eqref{eqthmJE} can also be interpreted as saying that this angular velocity has to be smaller than unit (or that the mean velocity is smaller than $c$). Thus, indeed seems reasonable to associate this inequality with the causal structure of the theory.

In order to study the connection between causality and inequality~\eqref{eqthmJE} we shall analyse if DEC is not only a sufficient but a necessary condition. In fact, we want to show the opposite, namely that a noncausal NLED can satisfy inequality \eqref{eqthmJE}. In particular, we examine an example of $\mathcal{L}(F)$ NLED, in which case, causality can be expressed as the condition~\citep{Goulart2009, Shabad2011}
\begin{equation} \label{eq:Scausal}
\mathcal{L}_{F}\le0 \ ,
\end{equation}
which together with unitarity is equivalent to imposing DEC. Thus, consider the lagrangian introduced by Kruglov in~\citep{Kruglov2017} as a modification of exponential electrodynamics
\begin{equation} \label{eq:Krug}
\mathcal{L}=-{F\over2}e^{-{F\over{2\beta^2}}} \ .
\end{equation}
Then, the causality condition reads
\begin{equation} \label{eq:KrugC}
B^2\le2\beta^2+E^2 \ ,
\end{equation}
which can be seen as an upper bound for the magnetic field with respect to the electric field. Since DEC guarantees the validity of inequality \eqref{eqthmJE}, we shall focus only on the noncausal configurations, i.e. $\mathcal{L}_{F}>0$. The difference between energy and angular momentum reads
\begin{align}
\mathcal{E}(\Sigma)-{\vert J(\Sigma)\vert\over\mathcal{R}}=&\int_{\Sigma}e^{-{F\over{2\beta^2}}}\left({E^2+B^2\over2}-{F\over{2\beta^2}}E^2 \right)\nonumber\\
&-{1\over\mathcal{R}}\left| \int_{\Sigma} e^{-{F\over{2\beta^2}}}\left({F\over{2\beta^2}}-1 \right)\epsilon_{ijk}\epsilon^{iab}B_{a}E_{b}k^{j}x^{k}\right| \ .
\end{align}

Following the same approach used in section~\ref{BekIneq:BIF}, the righthand side of the above expression can be majored as
\begin{align} \label{eq:fin}
\mathcal{E}(\Sigma)-{\vert J(\Sigma)\vert\over\mathcal{R}}\ge\int_{\Sigma}e^{-{F\over{2\beta^2}}}&\left\{{E^2+B^2\over2}+{x\over\mathcal{R}}EB\vert\sin\theta\vert \right.\nonumber\\
&\left. -{F\over{2\beta^2}}\left(E^2+{x\over\mathcal{R}}EB\vert\sin\theta\vert\right)\right\} \ .
\end{align}

It is straightforward to check that certain field's configurations can, indeed, satisfy the inequality between energy and angular momentum. The crucial point is to check if certain field's configurations that violate the causal condition can simultaneously satisfy the inequality. 
\begin{figure}
\centering
\includegraphics[width=0.5\textwidth]{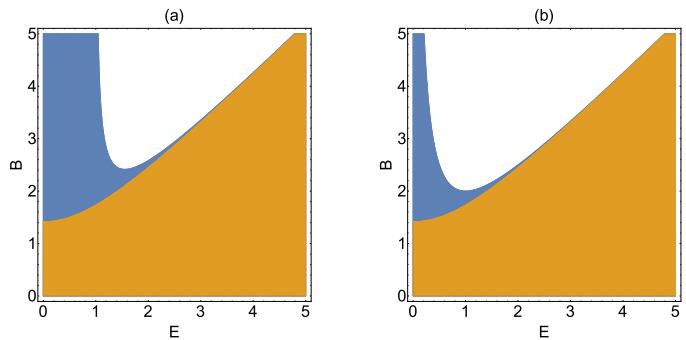}
\caption{The values where the integrand in \eqref{eq:fin} is positive for $x=\mathcal{R}$ (blue region) compared to the values where the causality condition \eqref{eq:KrugC} holds (orange region) for (a) $\theta=0$ and (b) $\theta=\pi/2$.}
\label{fig:DEC1}
\end{figure}
Fig \ref{fig:DEC1} shows that, indeed, there are field's configurations that violates the causal condition eq.~\eqref{eq:KrugC} but satisfy inequality~\eqref{eqthmJE}. Therefore, DEC is only a sufficient but not a necessary condition. In fact, one should be very careful to regard inequality~\eqref{eqthmJE} as a causal condition.

\section{Conclusion and Perspectives}\label{conclusion}

In the present work we investigated some physically motivated inequalities relating the charge, energy and angular momentum in the context of NLED. These inequalities are a direct consequence of the generalised Bekenstein-Mayo inequality eq.~\eqref{eq:BekMayo}. We have proved that, similarly to Maxwell theory, Born-Infeld electrodynamics satisfies all three inequalities but there is no rigidity statement inasmuch Born-Infeld total energy is always greater than Maxwell electromagnetic energy.

We have also shown that the inequality between charge and energy is independent to the quasi-local inequality relating energy and angular momentum by presenting a counter-example where only one of these inequalities is violated. Furthermore, this result suggest that these inequalities can be used, apart to the obvious Maxwell limit condition for weak fields, as a physically motivated criteria to select between NLED.

The fact that DEC is a sufficient condition to prove theorem~\ref{thmJE} indicates a possible relationship between the inequality between energy and angular-momentum and causality. Notwithstanding, we have shown that DEC is only a sufficient but not a necessary condition to this inequality, and hence, obscuring its physical content. It would be interesting to find a modified inequality that is a necessary and sufficient condition of DEC.


\begin{acknowledgments}
We would like to thank CNPq of Brazil for financial support.
\end{acknowledgments}

\appendix

\section{DEC and the Inequality between Energy and Angular Momentum}\label{app:DEC}

In this appendix we shall reproduce the argument of Dain~\cite{Dain2015} showing that the dominant energy condition is a sufficient condition to prove quasi-local inequality~\eqref{eqthmJE} between energy and angular-momentum.

Consider an arbitrary energy-momentum tensor $T_{\mu\nu}$ associated with some field theory. Given a timelike congruence $v^\mu$, the three-dimensional hypersurface $\mathcal{V}$ orthogonal to the congruence defined the its rest space. The energy associated with the observer's worldline is defined as
\begin{align}\label{app:defE}
\mathcal{E}=\int_{\mathcal{V}} T_{\mu\nu} v^\mu v^\nu\ .
\end{align}

If $\omega^\mu$ is a Killing vector field associated to space rotations, the angular-momentum can be defined as 
\begin{align}\label{app:defJ}
J(\mathcal{V})=\frac1c\int_{\mathcal{V}} T_{\mu\nu} v^\mu \omega^\nu\ .
\end{align}

Since the background is a flat Minkowski spacetime, we can choose Cartesian coordinates where $x^i$ expand $\mathcal{V}$ and $v^\mu=(1,0,0,0)$. In these coordinates, the rotation vector defined with respect to the direction $\hat{n}$ reads
\begin{align}\label{app:rotvec}
\omega_i=\epsilon_{ijk}n^jx^k\ .
\end{align}

The norm of the rotation vector reads $\omega\equiv \sqrt{-\omega^\mu \omega_\mu}=\sqrt{\omega_i \omega^i}$, hence we can define a spacelike unitary vector as $\hat{\omega}^\mu=\omega^\mu/\omega$.

The dominant energy condition implies that for all future-directed timelike $\xi^\mu$ or null $k^\mu$ vectors, the energy-momentum tensor satisfies 
\begin{align}\label{app:dec}
T_{\mu\nu}\, \xi^\mu k^\nu\geq0
\ .
\end{align}

In order to prove inequality~\eqref{eqthmJE} we  can choose a timelike vector $\xi^\mu=v^\mu$ and a null vector $k^\mu=v^\mu-\hat{\omega}^\mu$. From eq.~\eqref{app:dec} we have
\begin{align}\label{app:floowdec}
T_{\mu\nu}\, v^\mu v^\nu\geq T_{\mu\nu}\, v^\mu \hat{\omega}^\nu
\ .
\end{align}

The radius $\mathcal{R}$ of the minimum sphere $\mathcal{B_R}$ (definition~\ref{defR}) encloses all region $\Sigma$, and hence by definition we have $\omega \leq \mathcal{R}$. Therefore, we have
\begin{align}\label{app:proof}
\mathcal{E}(\Sigma)&=\int_{\Sigma} T_{\mu\nu} v^\mu v^\nu \nonumber\\
&\geq \int_{\Sigma} T_{\mu\nu} v^\mu \hat{\omega}^\nu\nonumber\\
&\geq \frac{1}{\mathcal{R}}\int_{\Sigma} T_{\mu\nu} v^\mu \omega^\nu
=\frac{c J(\Sigma)}{\mathcal{R}} \ .
\end{align}


\end{document}